\pdfoutput=1
\documentclass[journal]{IEEEtran}
\usepackage{graphicx}
\usepackage{adjustbox}
\usepackage{amsmath}  
\usepackage{cite}
\usepackage{algorithmic}
\usepackage{array}
\usepackage{dblfloatfix}
\usepackage{float}
\usepackage{amssymb}
\usepackage{multirow}
\usepackage{tikz}
\usepackage{tabularx}
\usepackage{booktabs}
\usepackage{adjustbox}
\usepackage{balance}
\usepackage{placeins}
\usepackage{afterpage} 
\usepackage{makecell}
\usepackage[symbol]{footmisc} 
\usepackage{enumitem}

\usepackage{xcolor}
\usepackage{soul}
\sethlcolor{yellow}

\usepackage{algorithm}
\usepackage{algorithmic}

\usepackage{url}

\usepackage{fancyhdr}

\ifCLASSINFOpdf
\else
\fi

\usepackage{amsmath} 

\fancypagestyle{ieee}{ 
\fancyhf{} 
\fancyhead[C]{
\footnotesize This article has been accepted for publication in IEEE Transactions on Circuits and Systems–I: Regular Papers. This is the author’s version which has not been fully edited and content may change prior to final publication. Citation information: DOI: 10.1109/TCSI.2025.3591960}
\fancyfoot[C]{
\footnotesize \copyright~2025 IEEE. Personal use of this material is permitted.  Permission from IEEE must be obtained for all other uses, in any current or future media, including reprinting/republishing this material for advertising or promotional purposes, creating new collective works, for resale or redistribution to servers or lists, or reuse of any copyrighted component of this work in other works.}

}
\begin{document}

\title{DiP: A Scalable, Energy-Efficient Systolic Array for Matrix Multiplication Acceleration}

\author{Ahmed~J.~Abdelmaksoud,
        Shady~Agwa,
        Themis~Prodromakis
\thanks{This work was supported in part by the Engineering and Physical Sciences Research Council (EPSRC) Programme Grant “Functional Oxide Reconfigurable Technologies” (FORTE) under Grant EP/R024642/2 and in part by the RAEng Chair in Emerging Technologies under Grant CiET1819/2/93.}

\thanks{Ahmed J.Abdelmaksoud, S. Agwa and T. Prodromakis are with the Centre for Electronics
Frontiers, Institute for Integrated Micro and Nano Systems, School of Engineering, The University of Edinburgh, EH9 3BF, Edinburgh, United Kingdom
(e-mails: a.j.abdelmaksoud@ed.ac.uk; shady.agwa@ed.ac.uk; t.prodromakis@ed.ac.uk).}}%

\maketitle

\thispagestyle{ieee}

\begin{abstract}
Transformers are gaining increasing attention across Natural Language Processing (NLP) application domains due to their outstanding accuracy. However, these data-intensive models add significant performance demands to the existing computing architectures. Systolic array architectures, adopted by commercial AI computing platforms like Google TPUs, offer energy-efficient data reuse but face throughput and energy penalties due to input-output synchronization via First-In-First-Out (FIFO) buffers.

This paper proposes a novel scalable systolic array architecture featuring \b{D}iagonal-\b{I}nput and \b{P}ermutated weight stationary (DiP) dataflow for matrix multiplication acceleration. The proposed architecture eliminates the synchronization FIFOs required by state-of-the-art weight stationary systolic arrays. Beyond the area, power, and energy savings achieved by eliminating these FIFOs, DiP architecture maximizes the computational resource utilization, achieving up to 50\% throughput improvement over conventional weight stationary architectures. Analytical models are developed for both weight stationary and DiP architectures, including latency, throughput, time to full PEs utilization (TFPU), and FIFOs overhead. 
A comprehensive hardware design space exploration using 22nm commercial technology demonstrates DiP's scalability advantages, achieving up to a 2.02× improvement in energy efficiency per area. Furthermore, DiP outperforms TPU-like architectures on transformer workloads from widely-used models, delivering energy improvement up to 1.81× and latency improvement up to 1.49×. At a 64×64 size with 4096 PEs, DiP achieves a peak throughput of 8.192 TOPS with energy efficiency 9.548 TOPS/W.

\end{abstract}


\IEEEpeerreviewmaketitle

\section{Introduction}

\IEEEPARstart{A}{rtificial} Intelligence (AI) is increasingly dominating various application domains \cite{1}. Natural Language Processing (NLP) is one of the crucial AI application domains receiving increasing attention nowadays \cite{2,3}. Thanks to Transformers, NLP tasks have been revolutionized by providing highly effective and scalable models for language understanding and generation \cite{4}. The exceptional capabilities of these models have led to a transformer-driven transformation across numerous application domains, including Machine Translation \cite{5}, Speech Recognition \cite{6}, Multimodal Applications \cite{7}, and Computer Vision \cite{8}.

However, transformers are data-intensive models that handle massive workloads compared to Deep Neural Networks (DNNs) and Convolutional Neural Networks (CNNs) \cite{9}. Additionally, transformer models have been growing exponentially, evolving from the original vanilla Transformer model with around 65 million parameters to models with hundreds of billions of parameters \cite{10,11}. A clear example of the challenging level of scalability is GPT (Generative Pre-trained Transformer) model that incorporates billions of parameters, primarily involving matrix multiplications \cite{12}.

Conventional Von-Neumann architectures are struggling to meet these increasing performance demands due to the memory/data-movement bottleneck. Systolic arrays, introduced in 1982, are spatial architectures that designed to maximize the data utilization to mitigate the memory bottleneck.
These architectures are receiving increased attention nowadays as a promising architecture for AI hardware acceleration \cite{13}. Typically, a systolic array consists of a two-dimensional (2D) interconnected Processing Elements (PEs). The PE consists of basic arithmetic, mainly multiplication and accumulation, along with register units. Systolic arrays are spatial architectures that enhance local data utilization by increasing the number of computation operations per each memory access. Accordingly, input data circulates among PEs in a wave-like fashion, while the communication with the synchronization First-In-First-Out (FIFO)s occurs only at the boundary PEs. Moreover, the interconnection of the systolic array naturally facilitates data reusability among PEs, particularly during matrix multiplications \cite{14}.

Although many systolic arrays are used for hardware acceleration \cite{15,16,17,18,19,20,21}, different adopted dataflows require additional synchronization hardware, which limits the energy efficiency and constrains performance capabilities. Therefore, transformers with massive matrix multiplication workloads will face significant challenges in leveraging the systolic arrays for the next generation of AI hardware.

Tensor Processing Unit (TPU) is a well-known AI computing architectures introduced by Google to handle massive matrix multiplication workloads with higher performance and energy efficiency than CPUs and GPUs \cite{22}. TPUs adopt weight stationary (WS) dataflow, which maximizes the data utilization of both weights and inputs. Google TPU V1 is designed primarily for inference, featuring a 256×256 systolic array optimized for 8-bit integer (INT8) operations, achieving a peak throughput of 92 TOPS \cite{23}. TPU V2 shifted to mixed-precision training with a smaller 128×128 systolic array per core optimized for FP16 and bfloat16 operations, boosting throughput to 180 TeraFLOPS \cite{24}. TPU V3 and V4 maintained the same array size but TPU V3 doubled the throughput to 420 TFLOPS per chip, aided by high memory bandwidth, while TPU v4 integrates four cores of 128×128 architecture, achieving up to 1 PFLOPS \cite{25}. 

Meissa is a systolic architecture that separates multipliers from adders instead of integrating them into a unified array \cite{26}. Similar to TPUs, it employs WS dataflow for the systolic array but eliminates input synchronization FIFOs to reduce overall latency. However, it has bulky adder trees per each column of the systolic array. These adder trees impose scalability limitations due to serious physical implementation challenges. As adder trees become larger, they require deeper pipelines to achieve higher frequencies. This increases the overall latency, area, and energy consumption. In addition, routing congestion presents another costly challenge, when all partial products from each PE column are delivered to the adder tree. Consequently, Meissa is not scalable to large N×N dimensions, which is critical for computation-intensive workload. Moreover, it still requires the output synchronization FIFOs, which add additional area, power, and energy penalty.

In this paper, we present a novel Diagonal-Input Permutated (DiP) weight stationary systolic array that overcomes the main challenges of the conventional WS systolic array architectures. The main contributions of this work are highlighted as follows:
\begin{itemize}
  \item We introduce DiP, a novel scalable spatial architecture of N×N PEs for matrix multiplication acceleration. DiP maximizes PE utilization and energy efficiency, achieving a throughput improvement of up to 1.49×.
  \item The proposed architecture eliminates the input/output synchronization FIFOs in WS by implementing a new dataflow with diagonal-input movement and permutated weights.
  \item The analytical models are extracted for DiP and WS including throughput, latency, time to full PE utilization (TFPU), and register overhead for different systolic array sizes.
  \item Hardware design space exploration and implementation are presented for DiP and WS using commercial 22nm technology, offering up to 2.02× energy efficiency per area improvement.
  \item DiP is evaluated using various transformer workloads from widely used models, outperforming TPU-like architectures and achieving up to 1.81× energy efficiency improvement and up to 1.49× latency improvement.
\end{itemize}

This paper is organized as follows; Section II discusses the systolic arrays background. Section III presents DiP architecture. Section IV shows hardware design space exploration, evaluation, and results. Finally, section V concludes the work.

\section{Systolic Arrays Background}

Systolic arrays adopt one of the dataflows to control the data movement across the PEs. Each dataflow retains one of the input, output, or weight data to be stationary during computations for the maximum duration to maximize the data reuse \cite{27,28}. The following dataflows are the most common in systolic array design: 
\begin{itemize}
  \item Weight Stationary (WS): weights are initially loaded to the PEs and kept stationary during processing. The input matrix and partial summations move among the PEs.
  \item Input Stationary (IS): input matrix is initially loaded to the systolic array, while weight matrix and partial summations move among the PEs.
  \item Output Stationary (OS): input and weight matrices move among the PEs, while the partial summations are accumulated inside PEs.
  \item Row Stationary (RS): it is proposed by Eyeriss \cite{29}. It adopts spatial architecture that uses coarse-grained PEs with internal memories to store weights and inputs. Inputs are broadcasted diagonally across the PEs, while weight matrix broadcasted horizontally, and partial summations move vertically.    
\end{itemize}

The OS dataflow moves both input and weight matrices simultaneously, which effectively doubles the required memory bandwidth for the systolic array. In RS dataflow, data redundancy increases because copies of the data are loaded into multiple PEs. Additionally, the circulation of weights within each PE reduces energy efficiency. On the other hand, WS dataflow is widely used in many architectures, such as Google TPU, due to its scalability and flexibility in handling convolutions and matrix multiplication \cite{23,24,25}. Moreover, it requires less memory bandwidth. Therefore, we focus on improving this dataflow.

\subsection{WS Dataflow}
\begin{figure}[tb]
  \centering
  \includegraphics[width=\linewidth, keepaspectratio]{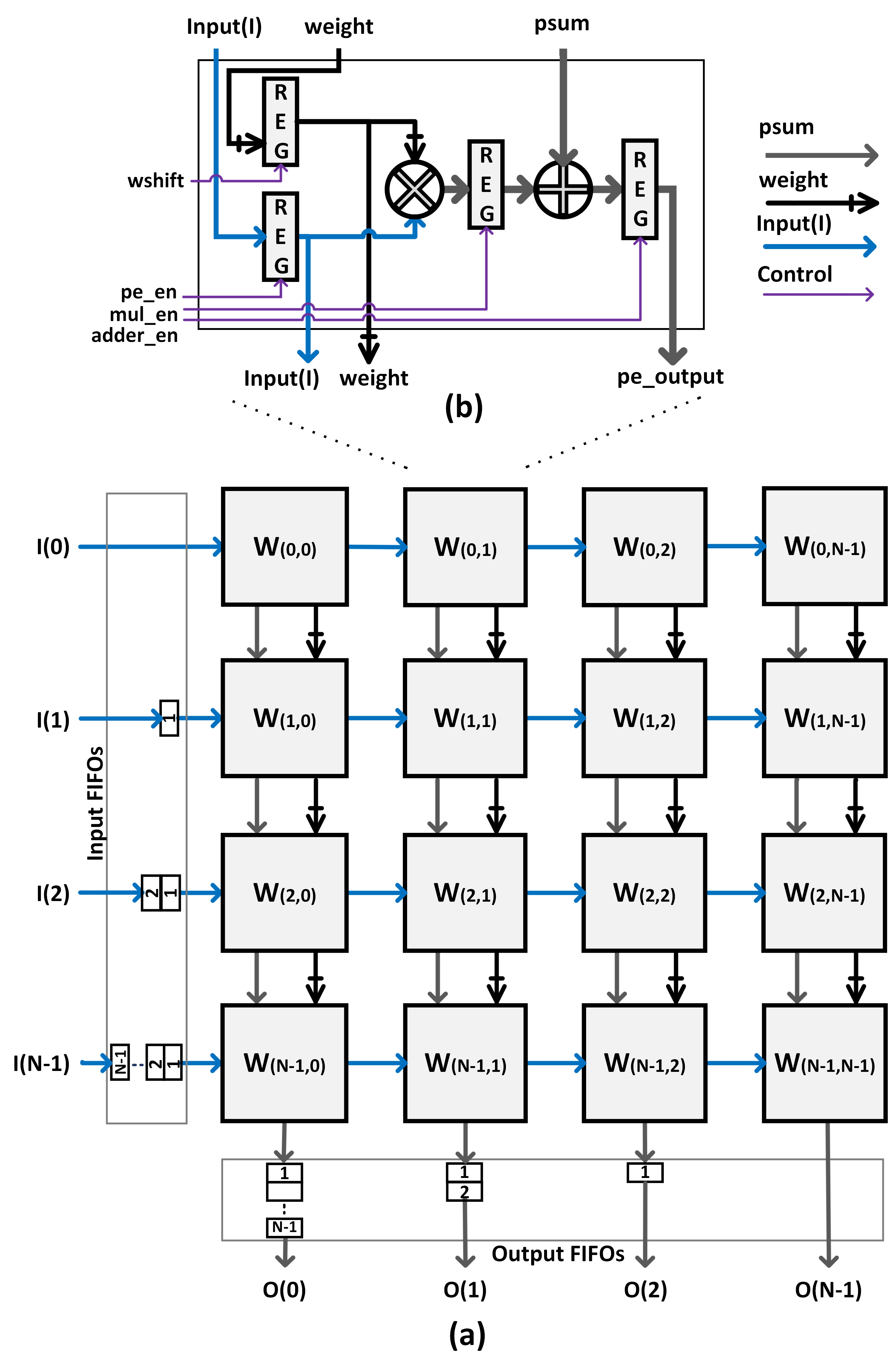} 
  \caption{(a) Top-level schematic for the WS systolic array with N×N PEs and two input/output FIFO groups. The weights (black crossed buses) are loaded vertically, and psums (grey buses) are accumulated along the columns. (b) WS PE block diagram, consisting of a 2-stage pipelined MAC unit and four enabled registers. Control signals \textit{wshift}, \textit{pe\_en}, \textit{mul\_en}, and \textit{adder\_en} are used for operations control. \textit{wshift} is shared across each PE row, while \textit{pe\_en}, \textit{mul\_en}, and \textit{adder\_en} are shared across each PE column.}
  \label{Fig.1}
\end{figure}

The WS dataflow is widely used in many architectures where weights are initially loaded to the PEs, while the input matrix circulates between them in a systolic fashion.
This approach mitigates the memory bottleneck by reducing memory accesses and increasing data reuse. However, WS dataflow requires input and output FIFOs to synchronize data for proper functionality. 
Figure \ref{Fig.1} shows the architecture of WS systolic array consisting of N×N PEs and two FIFO groups for input/output synchronization. 
WS PE consists of 2-stage pipelined MAC unit and four enabled registers for input, weight, multiplier output and adder output. 
Control signals \textit{wshift}, \textit{pe\_en}, \textit{mul\_en}, and \textit{adder\_en} manage the PE operations. Specifically, \textit{wshift} enables the weight register, while \textit{pe\_en} enables the input register. \textit{mul\_en} and \textit{adder\_en} selectively enable their respective registers only during active computation cycles, reducing power consumption during inactive cycles.
\textit{wshift} is shared across each PE row, while \textit{pe\_en}, \textit{mul\_en}, and \textit{adder\_en} are shared across each PE column.
The input FIFO group consists of series of FIFOs with incrementally increasing depth starting from one element in the second row to $N-1$ elements in the last row. 
The output FIFO group is structured similarly to the input FIFO group, but the FIFO depths decrease from $N-1$ elements to one element moving from left to right across the systolic array columns. 
 
The input FIFOs map the input matrix to the PEs in diagonal wave fashion, while the output FIFOs synchronize partial summations along each column \cite{27}. In WS, the input matrix and partial summations (psums) are pipelined to ensure proper processing and improve data reuse. The PE in WS is fine-grained, unlike the coarse-grained PE in Eyeriss, which broadcasts weights and input matrices to the array and stores them in internal memories \cite{29}. Weights are loaded vertically, while the input matrix is shifted horizontally, while the output is shifted to the output FIFO group. Using synchronization FIFOs add significant overheads in terms of area, power, energy consumption, and latency during matrix multiplication. Moreover, the computations propagate as a diagonal wave from the top-left corner to the bottom-right corner of the systolic array. This substantially reduces overall PE utilization resulting in degraded performance and increased latency. As a result, FIFO-based systolic arrays suffer from lower energy efficiency, reduced PE utilization, higher latency, and decreased throughput.

The analytical modeling of the WS systolic array is studied for latency, throughput, TFPU, and FIFOs overhead, providing insights into dataflow performance. For the WS latency analytical model, as shown in \eqref{eq:1}, WS requires $3N+S-3$ cycles to complete the processing, where $N$ represents the number of rows/columns per WS systolic array, and $S$ is the number of pipeline stages per Multiply-Accumulate (MAC) unit, which equals one for 1-stage pipelined MAC, and two for 2-stage pipelined MAC. 
The throughput, as indicated in \eqref{eq:2}, is calculated as the ratio of total operations to WS latency. 
Regarding register overhead, WS systolic array uses two FIFO groups for input and output synchronization. Each group consists of $N-1$ FIFOs, as shown in Fig. \ref{Fig.1}. Additionally, each FIFO group includes ${N(N-1)/2}$ registers. Consequently, the total register overhead for a typical WS systolic array is calculated as shown in \eqref{eq:3}.
TFPU is another metric introduced to calculate the required number of cycles to reach full utilization of PEs. This metric shows the overhead when the input matrix is initially loaded to the PEs. TFPU is calculated, as shown in \eqref{eq:4}, where it takes $2N-1$ cycles for WS to reach full PEs utilization.

\begin{equation}
    \text{Latency for WS} = 3N + S - 3 \tag{1}\label{eq:1}
\end{equation}
\begin{equation}
    \text{Throughput for WS} = \frac{2N^3}{3N + S - 3} \tag{2}\label{eq:2}
\end{equation}
\begin{equation}
    \text{Registers overhead for WS} = N(N - 1) \tag{3}\label{eq:3}
\end{equation}
\begin{equation}
    \text{TFPU for WS} = 2N - 1 \tag{4}\label{eq:4}
\end{equation}

\section{DiP Architecture}

\begin{figure*}[tb] 
  \centering
  \includegraphics[width=\textwidth, keepaspectratio]{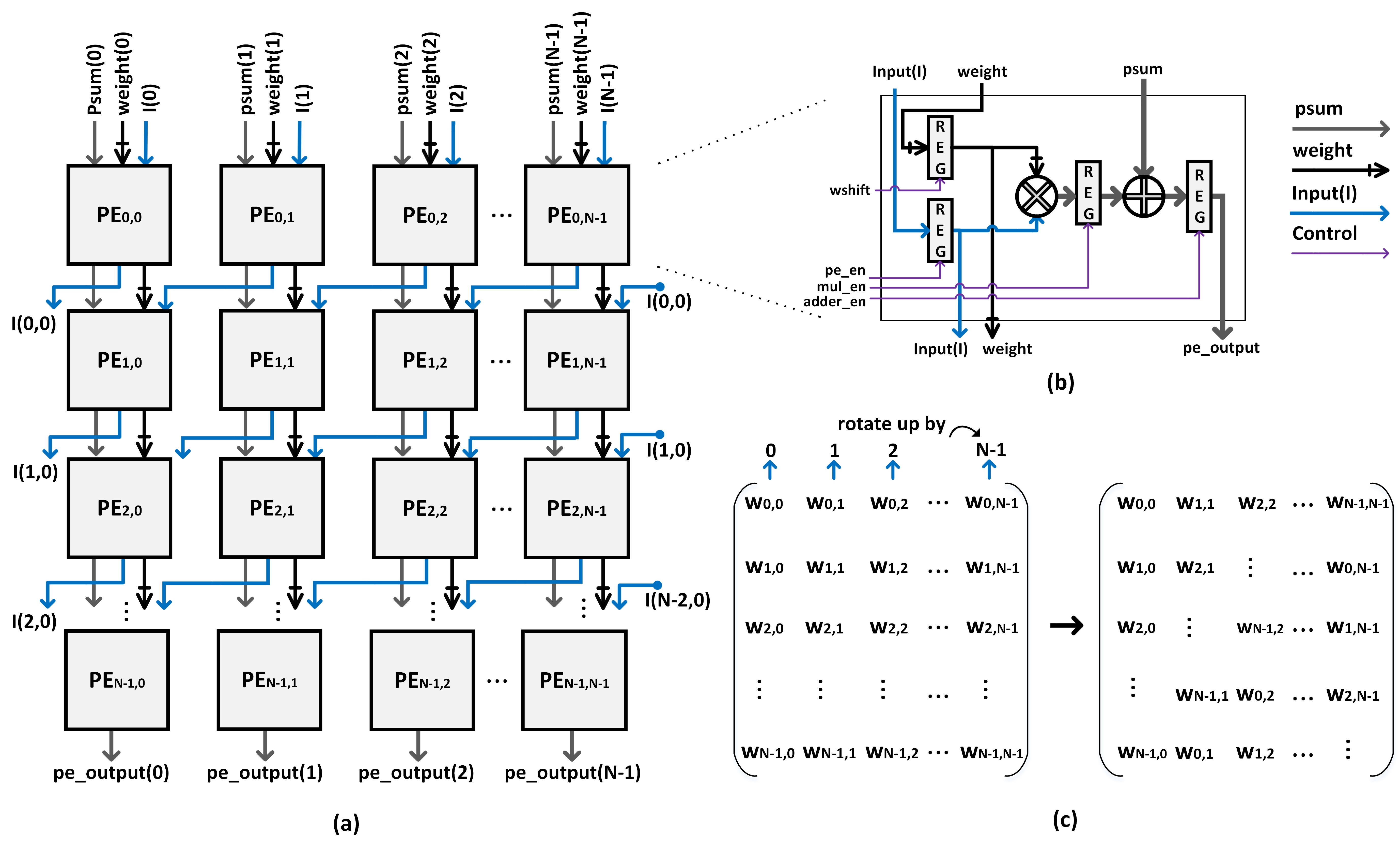} 
  \caption{(a) N×N DiP systolic array architecture, inputs(I) move diagonally across PE rows, transitioning from one row to the next. The boundary PEs are diagonally connected, so that the registered inputs from the leftmost PE column feed into the inputs of the rightmost PE column in the subsequent row. Weights are loaded vertically, and psums are accumulated vertically along the columns as well. (b) PE block diagram, consisting of a 2-stage pipelined MAC unit and four enabled registers. Control signals \textit{wshift}, \textit{pe\_en}, \textit{mul\_en}, and \textit{adder\_en} are used for operations control. \textit{wshift} is shared between all systolic array PEs, while \textit{pe\_en}, \textit{mul\_en}, and \textit{adder\_en} are shared across each PE row. (c) General weight matrix permutation for DiP dataflow. The weight matrix is permutated by shifting and rotating each column by its column index}
  \label{Fig.2}
\end{figure*}

\begin{figure*}[h] 
  \centering
  \includegraphics[width=\textwidth, keepaspectratio]{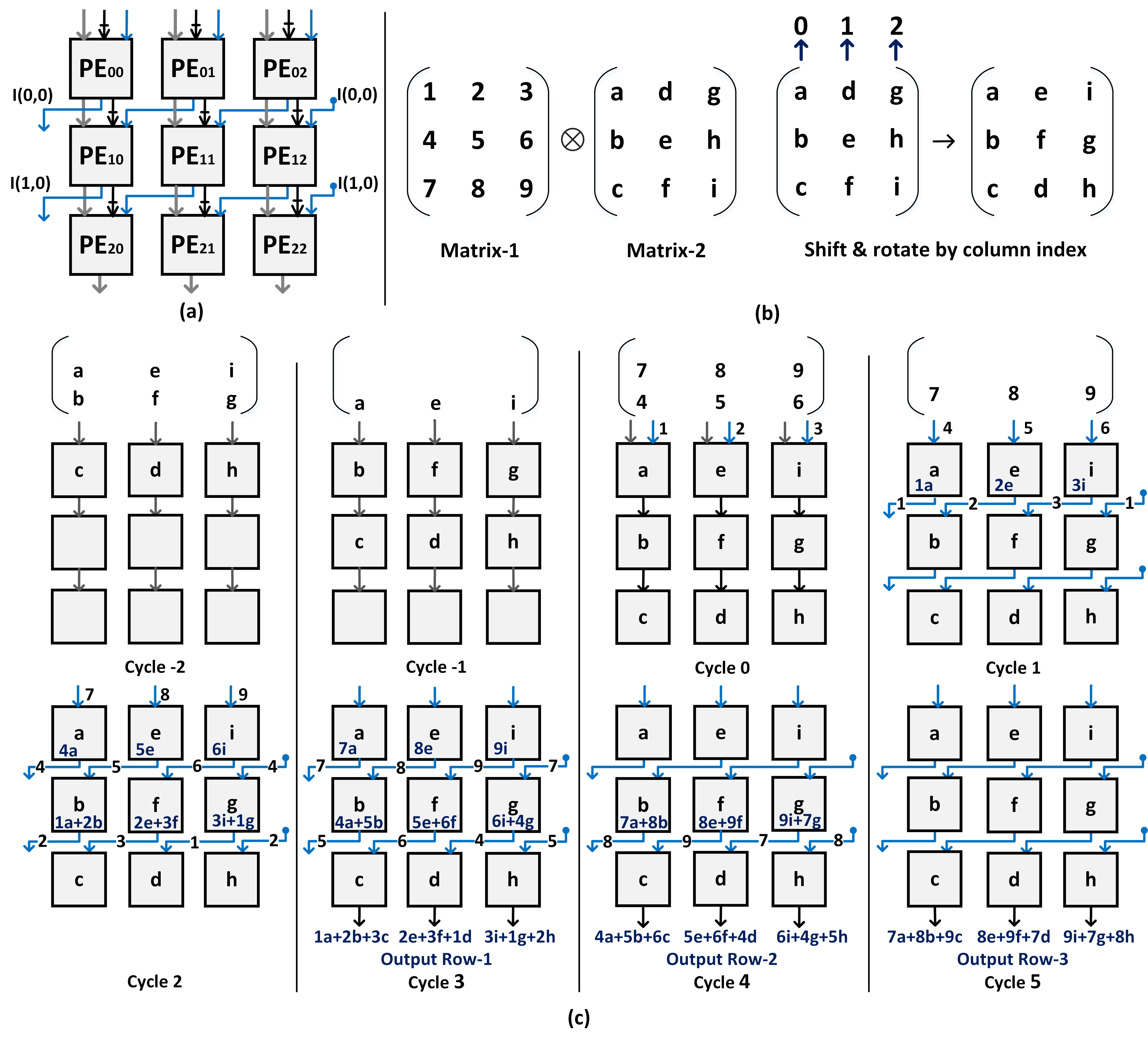} 
  \caption{Example for 3×3 DiP systolic array (a) shows the diagonal input connections for 3×3 DiP, (b) shows 3×3 DiP weight matrix permutation by shifting and rotating each column by its column index, and (c) shows the processing flow for 3×3 DiP example cycle by cycle. \textit{Cycles (-2, -1, 0)} are dedicated to weight matrix loading, while \textit{Cycle 0} involves loading the last row of the weight matrix and the first row of the input matrix. Cycles from \textit{Cycle 1} to \textit{Cycle 5} are allocated for matrix multiplication processing, while the first output row becomes ready at \textit{Cycle 3}.}
  \label{Fig.3}
\end{figure*}

 In this section, we discuss DiP architecture, DiP dataflow, and DiP analytical models. Then, we compare the analytical models for DiP and WS.

 \subsection{DiP Architecture}
 DiP is a scalable spatial architecture consisting of N×N PEs, designed to accelerate matrix multiplication computations, as shown in Fig. \ref{Fig.2}(a). The input matrix moves diagonally across the PEs, passing from one PE row to the next. The boundary PEs are diagonally connected, with the registered inputs of the leftmost PE column are connected to the inputs of the rightmost PE column in the subsequent row. The weight matrix is loaded vertically, and psums are accumulated vertically. Figure \ref{Fig.2}(b) shows the architecture of each PE. 
 
 The proposed PE is similar to WS PE, utilizing a 2-stage pipelined MAC unit and four enabled registers for input, weight, multiplier output and adder output. Control signals \textit{wshift}, \textit{pe\_en}, \textit{mul\_en}, and \textit{adder\_en} manage the PE operations. Specifically, \textit{wshift} enables the weight register, while \textit{pe\_en} enables the input register. \textit{mul\_en} and \textit{adder\_en} selectively enable their respective registers only during active computation cycles, reducing power consumption during inactive cycles. However in DiP, \textit{wshift}, \textit{pe\_en}, \textit{mul\_en}, and \textit{adder\_en} are shared across each PE row.

\subsection{DiP Dataflow}
The proposed DiP architecture adopts a novel dataflow to control the movement of inputs, weights, and partial summations across the whole systolic array. DiP dataflow relies on two major upgrades compared to WS dataflow, the diagonal movement of input matrix, and the weight matrix permutation. Firstly, the input matrix moves diagonally across PE rows, as shown in Fig. \ref{Fig.2}(a). Secondly, the proposed dataflow permutates the weights by shifting and rotating each column by its column index, as shown in Fig. \ref{Fig.2}(c). The weights permutation is initially prepared on software level, as shown in Algorithm \ref{alg:matrix_permutation}. For each column, it iterates over all rows, assigning each element in the \textit{permutated\_matrix} based on its row and column index. The permutation is performed at the software level when the second matrix is a weight matrix. Alternatively, it is executed at run-time by efficiently re-scheduling memory access across multi-bank memories with almost zero overhead. The proposed dataflow removes the need for input and output synchronization FIFOs typically required by conventional WS systolic arrays. Additionally, it enhances throughput and PE utilization while reducing chip area and latency.

\begin{algorithm}
\caption{: Matrix Permutation for DiP Dataflow}
\label{alg:matrix_permutation}
\begin{algorithmic}[0]  
\STATE \textbf{Input}: $matrix$
\STATE \textbf{Output}: $permutated\_matrix$
\FOR{$i \gets 0$ to cols}
\FOR{$j \gets 0$ to rows}
    \STATE $permutated\_matrix[j][i] = matrix[(j + i) \% rows][i]$
\ENDFOR
\ENDFOR
\end{algorithmic}
\end{algorithm}

Figure \ref{Fig.3} shows a complete example for 3×3 DiP systolic array. It consists of three rows: a) Row-0: $PE_{00}$, $PE_{01}$, $PE_{02}$, b) Row-1: $PE_{10}$, $PE_{11}$, $PE_{12}$, c) Row-2: $PE_{20}$, $PE_{21}$, $PE_{22}$. The PE array is diagonally connected, as shown in Fig. \ref{Fig.3}(a). The leftmost PE in each PE row is connected to the rightmost PE in the next PE row. The weight matrix is initially permutated to be prepared for weights loading. Each column is shifted and rotated by its column index, as shown in Fig. \ref{Fig.3}(b). The weights are initially loaded, row by row, to the systolic array, as shown in Fig. \ref{Fig.3} from \textit{Cycle -2} to \textit{Cycle 0}. The loading of the last weight row and the first input matrix are performed in parallel at \textit{Cycle 0}. The input data is loaded in parallel to Row-0 including inputs for $PE_{00}$, $PE_{01}$, and $PE_{02}$. After accomplishing the required computations by Row-0, the input data is shifted diagonally from $PE_{00}$ to $PE_{12}$ and from $PE_{01}$ to $PE_{10}$ and from $PE_{02}$ to $PE_{11}$. Then after accomplishing the required computations by Row-1, the input data is shifted diagonally again from $PE_{10}$ to $PE_{22}$ and from $PE_{11}$ to $PE_{20}$ and from $PE_{12}$ to $PE_{21}$. The processing starts from \textit{Cycle 1} to \textit{Cycle 5}, while the first output row becomes ready at \textit{Cycle 3}, and the last output row becomes ready at \textit{Cycle 5}. The processing goes as follows:
\begin{itemize}
  \item \textbf{Cycle -2}: The last row of weight matrix \textit{(c, d, h)} is loaded to the first PE row. 
  \item \textbf{Cycle -1}: The weights matrix row \textit{(c, d, h)} is shifted to the second PE row, and new weights row \textit{(b, f, g)} is loaded to the first PE row. 
  \item \textbf{Cycle 0}: The weight matrix row \textit{(c, d, h)} is shifted to the last PE row, the weights in the first PE row \textit{(b, f, g)} are shifted to the second PE row, and new weights row \textit{(a, e, i)}  and the first input matrix row \textit{(1, 2, 3)} are loaded to the first PE row, simultaneously.
  \item \textbf{Cycle 1}: The first PE row shifts the partial summations \textit{(1a, 2e, 3i)} to the second row. Using the diagonal connections, the input matrix row \textit{(1, 2, 3)} is permutated to \textit{(2, 3, 1)}, and loaded to the second PE row at the same cycle. 
  \item \textbf{Cycle 2}: The first PE row shifts the partial summations \textit{(4a, 5e, 6i)} to the second row, and the input matrix row \textit{(4, 5, 6)} is permutated to \textit{(5, 6, 4)} and loaded to the second PE row. Similarly, The second PE row shift the partial summations \textit{(1a+2b, 2e+3f, 3i+1g)} to the third row, and the input matrix row \textit{(2, 3, 1)} is permutated to \textit{(3, 1, 2)} and loaded to the third PE row.
  \item \textbf{Cycle 3}: The first PE row shift the partial summations \textit{(7a, 8e, 9i)} to the second row, and the input matrix row \textit{(7, 8, 9)} is permutated to \textit{(8, 9, 7)} and loaded to the second PE row. Similarly, The second PE row shift the partial summations \textit{(4a+5b, 5e+6f, 6i+4g)} to the third row, and the input matrix row \textit{(5, 6, 4)} is permutated to \textit{(6, 4, 5)} and loaded to the third PE row. In addition, The first output row \textit{(1a+2b+3c, 2e+3f+1d, 3i+1g+2h)} is shifted out.
  \item \textbf{Cycle 4}: The first PE row becomes idle, unless more input rows are loaded, and the partial summations \textit{(7a+8b, 8e+9f, 9i+7g)} are shifted from the second PE row to the third row. In addition, the input matrix row \textit{(8, 9, 7)} is permutated to \textit{(9, 7, 8)}, and loaded to the third PE row. In addition, the second output row \textit{(4a+5b+6c, 5e+6f+4d, 6i+4g+5h)} is shifted out.
  \item \textbf{Cycle 5}: The first and second PE rows are idle, and the third output row \textit{(7a+8b+9c, 8e+9f+7d, 9i+7g+8h)} is shifted out.
\end{itemize}

\noindent Meanwhile, additional associated input tiles to the loaded weight tile are scheduled for larger input matrix, and the processing continues till the end of the workload.

\begin{figure*}[h] 
  \centering
  \includegraphics[width=\textwidth, keepaspectratio]{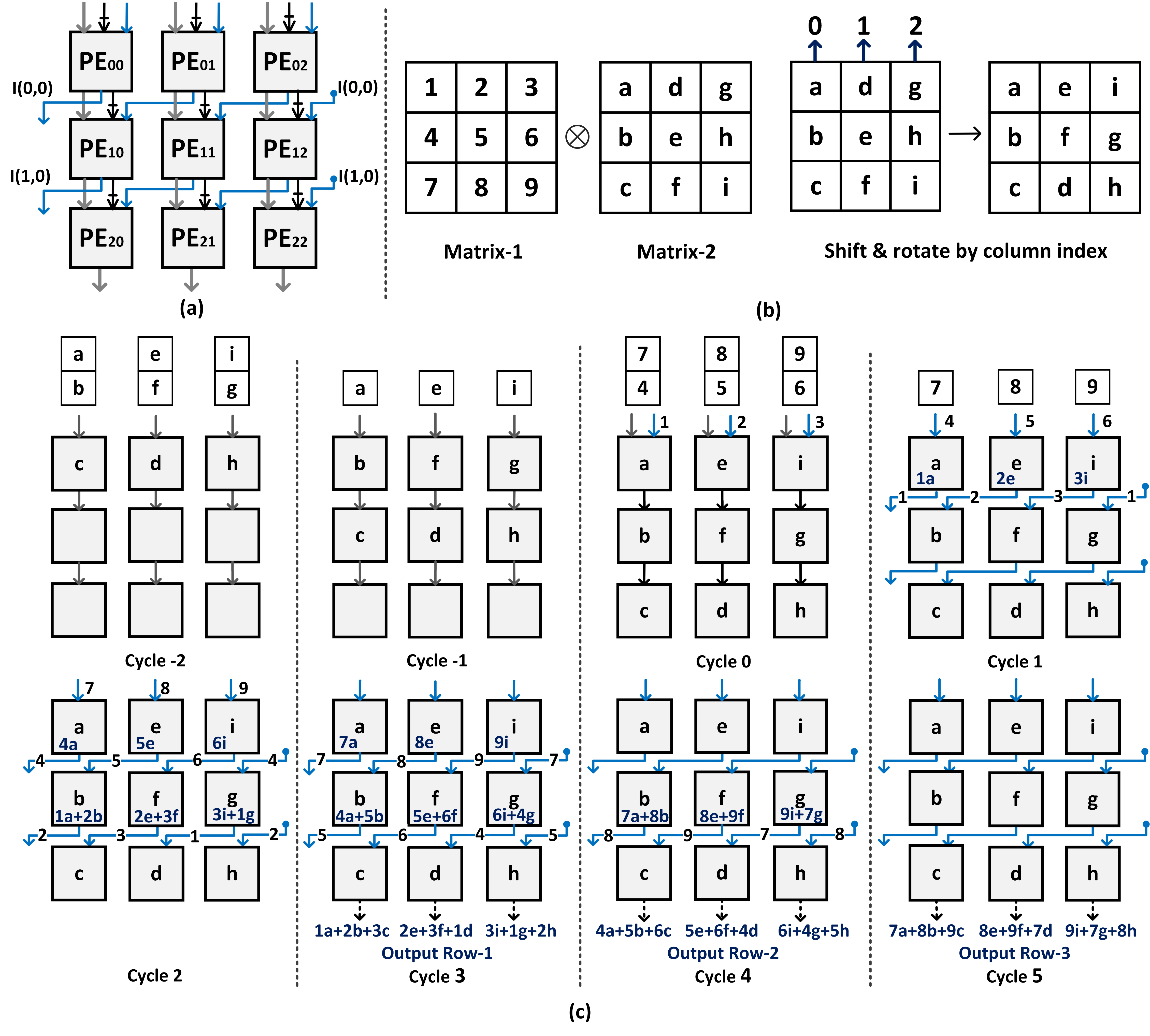} 
  \caption{Analytical modeling results comparing WS and DiP in terms of latency, throughput, register usage, and TFPU. (a) Latency (per single tile processing) for WS and DiP systolic arrays, with the grey curve indicating the percentage of latency savings of DiP over WS. (b) Throughput, measured in operations per cycle (OPS/Cycle), for WS and DiP, with the grey curve showing the percentage throughput improvement of DiP compared to WS. (c) Register usage for DiP compared to WS, normalized to 8-bit (baseline datawidth), with the grey curve indicating the percentage of register savings. (d) TFPU, representing the time required to activate all PEs, with the grey curve showing the percentage TFPU improvement. WS and DiP are represented in blue and black, respectively.}
  \label{Fig.4}
\end{figure*}

\subsection{DiP Analytical Model}
The analytical models for DiP are studied for latency and throughput. Regarding DiP latency, DiP systolic array requires \textit{2N+S-2} cycles for processing, where \textit{N} is the number of rows/columns per DiP systolic array, and \textit{S} is the number of pipelined stages per MAC unit. 
As a result, the latency is \textit{2N-1} cycles for 1-stage pipelined PE, and \textit{2N} cycles for 2-stage pipelined PE, as shown in \eqref{eq:5}. 
The throughput is calculated as the number of operations (multiplications and additions) divided by the latency, as shown in \eqref{eq:6}. 
TFPU is calculated, as shown in \eqref{eq:7}, where it takes $N$ cycles to reach full PEs utilization, outperforming WS by $N-1$ cycles.
Moreover, The registers overhead is eliminated by removing the input/output synchronization FIFOs. 

\begin{equation}
    Latency \ for \ DiP = 2N+S-2 \tag{5}\label{eq:5}
\end{equation}
\begin{equation}
    Throughput \ for \ DiP = \dfrac{2N^3}{2N+S-2} \tag{6}\label{eq:6}
\end{equation}
\begin{equation}
    TFPU \ for \ DiP = N \tag{7}\label{eq:7}
\end{equation}

\subsection{DiP/WSSA Analytical Comparison}
The scalability of systolic arrays is important to meet the acceleration requirements. The proposed systolic array is gradually scaled up from 3x3 to 64x64 with sizes (3x3, 4x4, 8x8, 16x16, 32x32, 64x64). An analytical comparison of DiP and WS is presented, evaluating throughput, latency, register savings, and TFPU across different systolic array sizes. Figure \ref{Fig.4}(a) shows the latency for DiP compared to WS, with the percentage of latency savings calculated as the difference between WS and DiP latencies, divided by WS latency. The saved percentage starts at 28\% for a 3x3 systolic array and reaches 33\% for a 64x64 systolic array.  

In addition, The throughput for both the DiP and WS systolic arrays is compared, as shown in Fig. \ref{Fig.4}(b). The throughput improvement is calculated as the ratio between DiP to WS throughput. The improvement ratio starts at 33.3\% for a 3x3 systolic array and reaches 49.2\% for a 64x64 systolic array.  The proposed architecture significantly increases the PEs utilization. Thus, it outperforms the conventional WS counterparts in terms of throughput by up to 50\%. 

Moreover, Fig. \ref{Fig.4}(c) shows the percentage of saved registers as another design improvement of DiP compared to the WS systolic array. By eliminating input/output FIFOs, the percentage of saved registers reaches up to 20\% for a 64×64 systolic array. The saved registers is calculated as the difference between WS and DiP used registers, divided by the number of registers used by WS. The registers of WS systolic array is distributed between input synchronization FIFOs, Output synchronization FIFOs, and internal PE registers. In contrast, the proposed diagonal input systolic array relies solely on internal PE registers, eliminating the need for input/output FIFOs. The represented number of registers are normalized to 8-bit as the baseline bandwidth.

TFPU calculates the required number of cycles to reach full utilization of PEs . This metric shows the overhead when the inputs are initially loaded, particularly for large matrix-matrix multiplication. Figure \ref{Fig.4}(d) shows TFPU for WS and DiP systolic arrays. The proposed DiP rapidly utilizes all PEs row by row, whereas WS gradually activates PEs in a diagonal pattern, starting from the top-left and moving to the bottom-right. Consequently, DiP outperforms WS, requiring only almost half the time of WS to fully utilize the entire systolic array.

\section{Evaluation \& Results}

This section explores the hardware design space for the proposed DiP architecture, followed by benchmarking DiP using transformers and evaluating it against TPU-like architecture across various transformer workloads. Finally, DiP is compared with existing accelerators in the literature.

\begin{table*}[bt]
\caption{Comparison of Area, Power Consumption, and Improvements in Throughput, Power Consumption, Area, and Overall Performance of DiP Compared to WS at Different Sizes using Commercial 22nm Technology at Frequency of 1GHz}

\label{table:area_power_saving}
\centering
\scriptsize 
\setlength{\tabcolsep}{6pt} 
\begin{tabular}{
>{\centering\arraybackslash}p{1.3cm} 
>{\centering\arraybackslash}p{1.0cm}   
>{\centering\arraybackslash}p{1.0cm}
>{\centering\arraybackslash}p{2cm}
>{\centering\arraybackslash}p{1.0cm}
>{\centering\arraybackslash}p{1.0cm}
>{\centering\arraybackslash}p{2.3cm}
>{\centering\arraybackslash}p{2cm}
>{\centering\arraybackslash}p{2.3cm}}
\hline
\textbf{Size} 
& \multicolumn{2}{c}{\textbf{Area (mm$^2$)}} 
& \textbf{Area Improvement ($\times$)} 
& \multicolumn{2}{c}{\textbf{Power Consumption (mW)}} 
& \textbf{Power Consumption Improvement ($\times$)} 
& \textbf{Throughput Improvement ($\times$)} 
& \textbf{Overall Improvement\textsuperscript{*} ($\times$)} 
\\
\cline{2-3} \cline{5-6}      
& \textbf{WS} 
& \textbf{DiP} 
& 
& \textbf{WS} 
& \textbf{DiP}  
& & \\ 
\hline
4×4   & 0.0052 & 0.0049 & 1.06 & 4.168 & 3.582 & 1.16 & 1.38 & 1.70 \\ 
8×8   & 0.0187 & 0.0174 & 1.08 & 16.2  & 13.72 & 1.18 & 1.44 & 1.84 \\ 
16×16 & 0.0712 & 0.0654 & 1.09 & 64.28 & 53.63 & 1.20 & 1.47 & 1.92 \\ 
32×32 & 0.2750  & 0.253  & 1.09 & 264.2 & 211.5 & 1.25 & 1.48 & 2.02 \\ 
64×64 & 1.085  & 1.012  & 1.07 & 1041  & 857.8 & 1.21 & 1.49 & 1.93 \\ 
\hline
\end{tabular}
\begin{flushleft}
\scriptsize \textsuperscript{*}Overall improvement represents energy efficiency per area.\\
\end{flushleft}
\end{table*}

\subsection{Hardware Design Space Exploration}

A hardware design space exploration is developed for DiP and WS at different sizes. Both designs are scaled from 4x4 to 64×64 with variants (4×4, 8×8, 16×16, 32×32, 64×64). A parameterized HDL design using Verilog is developed. Then, all designs are implemented from synthesize to GDSII using commercial 22nm technology at frequency of 1 GHz. Table \ref{table:area_power_saving} shows a comparison between WS and DiP at different sizes in terms of area and power consumption. Additionally, the improvements in throughput, power consumption, area, and overall improvement (energy efficiency per area) for different WS/DiP design space. DiP outperforms WS across all metrics, with overall performance from 1.7× to 2.02×. At a size of 32×32, DiP achieves 1.48× higher throughput than WS, 1.25× lower power consumption, and 1.09× smaller area footprint, resulting in a total improvement of 2.02×. Additionally, at size of 64×64, the throughput is improved by 1.49×, power consumption is reduced by 1.21×, and area is decreased by 1.07× compared to WS, with overall improvement 1.93×.

\subsection{Transformers Benchmarking}

Transformer workloads are becoming increasingly massive and heavily dependent on matrix multiplication, especially in Multi-Head Attention (MHA) and Feed-Forward Networks (FFN). MHA, a core component of transformer models proposed by Vaswani et al., enables the capture of complex dependencies in data \cite{10}. By leveraging multiple attention heads, MHA captures diverse representational subspaces, allowing the model to understand relationships across different perspectives simultaneously.

Table \ref{table:mha-ffn-workloads} presents the matrix multiplication workloads and their dimensions for MHA and FFN operations within transformer models, highlighting the relationship between input sequence length (\(l\)), model size (\(d_{\text{model}}\)), head size (\(d_k\)), and FFN size (\(d_{\text{FFN}}\)). The table offers insights into the computational requirements of transformer models, emphasizing how the sequence length and model size impact the matrix operations in both the attention mechanism and the feed-forward network. In MHA, the input (\( X \)) is first projected in Queries (\( Q_i \)), Keys (\( K_i \)), and Values (\( V_i \)) per each head \( i \) using learned weight matrices. The attention scores (\( S_i \)) for each head are computed by taking the scaled dot product of Queries (\( Q_i \)) and transposed Keys (\( K_i \)), followed by a softmax normalization. The attention scores (\( S_i \)) are multiplied with Values (\( V_i \)), producing the attention output (\( Attn_i \)) for each head. The outputs from all heads are concatenated into a single matrix \(\text{Attn}_\text{concat}\), which is finally projected back to the model's hidden dimension (\(d_{\text{model}}\)) using a learned output projection matrix (\( W^O \)). This process enables the model to capture information from multiple representation subspaces simultaneously, enhancing the model's ability to represent complex sequences effectively. FFN in transformers consists of two linear transformations with a non-linear activation function applied between them. The process begins with the first matrix multiplication, which projects the input \( Y \) into a higher-dimensional space using a weight matrix \( W_1 \) and adding bias \( b_1 \), as shown in Table \ref{table:mha-ffn-workloads}. Next, a non-linear activation function, such as ReLU or GELU is applied. Finally, the second matrix multiplication applies the second linear transformation to the non-linearly transformed output, mapping it back to the original dimensionality.

\begin{table}[bt]
\centering
\caption{Matrix Multiplication Dimensions for MHA and FFN Workloads in terms of Sequence Length (\(l\)), Model Hidden Size (\(d_{\text{model}}\)), Head Size (\(d_{k}\)), FFN Size (\(d_{\text{FFN}}\)). Input matrices Sizes are \(M \times N\) and \(N \times K\), while Output Matrix is \(M \times K\).}
\label{table:mha-ffn-workloads}
\resizebox{\linewidth}{!}{%
\begin{tabular}{llc}
\hline
\textbf{Workload} & 
\textbf{Dimensions} \\
& 
\textbf{(M, N, K)} \\
\hline
Input Projections, \([Q_i, K_i, V_i] = X \cdot [W^Q_i, W^K_i, W^V_i]\) 
& \(l \times d_{\text{model}} \times d_{k}\) \\
\(A_i = Q_i \cdot K_i^T\) 
&  \(l \times d_{k} \times l\) \\
Attention Scores, \(S_i = \text{Softmax}(A_i / \sqrt{d_k})\)
&  \(l \times d_{k} \times l\) \\
Attention, \(\text{Attn}_i = S_i \cdot V_i\) 
&  \(l \times l \times d_{k}\) \\
Output Projection, \(Y = \text{Concat}(\text{Attn}_1, \text{Attn}_2, \dots, \text{Attn}_h) \cdot W^O\) 
& \(l \times d_{\text{model}} \times d_{\text{model}}\) \\
\hline
FFN1 Projection, \(Z = \text{Non-Linear}(Y \cdot W_1 + b_1)\) 
& \(l \times d_{\text{model}} \times d_{\text{FFN}}\) \\
FFN2 Projection, \(F = Z \cdot W_2 + b_2\) 
& \(l \times d_{\text{FFN}} \times d_{\text{model}}\) \\
\hline
\end{tabular}
}
\end{table}

\begin{figure*}[!t] 
  \centering
  \includegraphics[width=\textwidth, keepaspectratio]{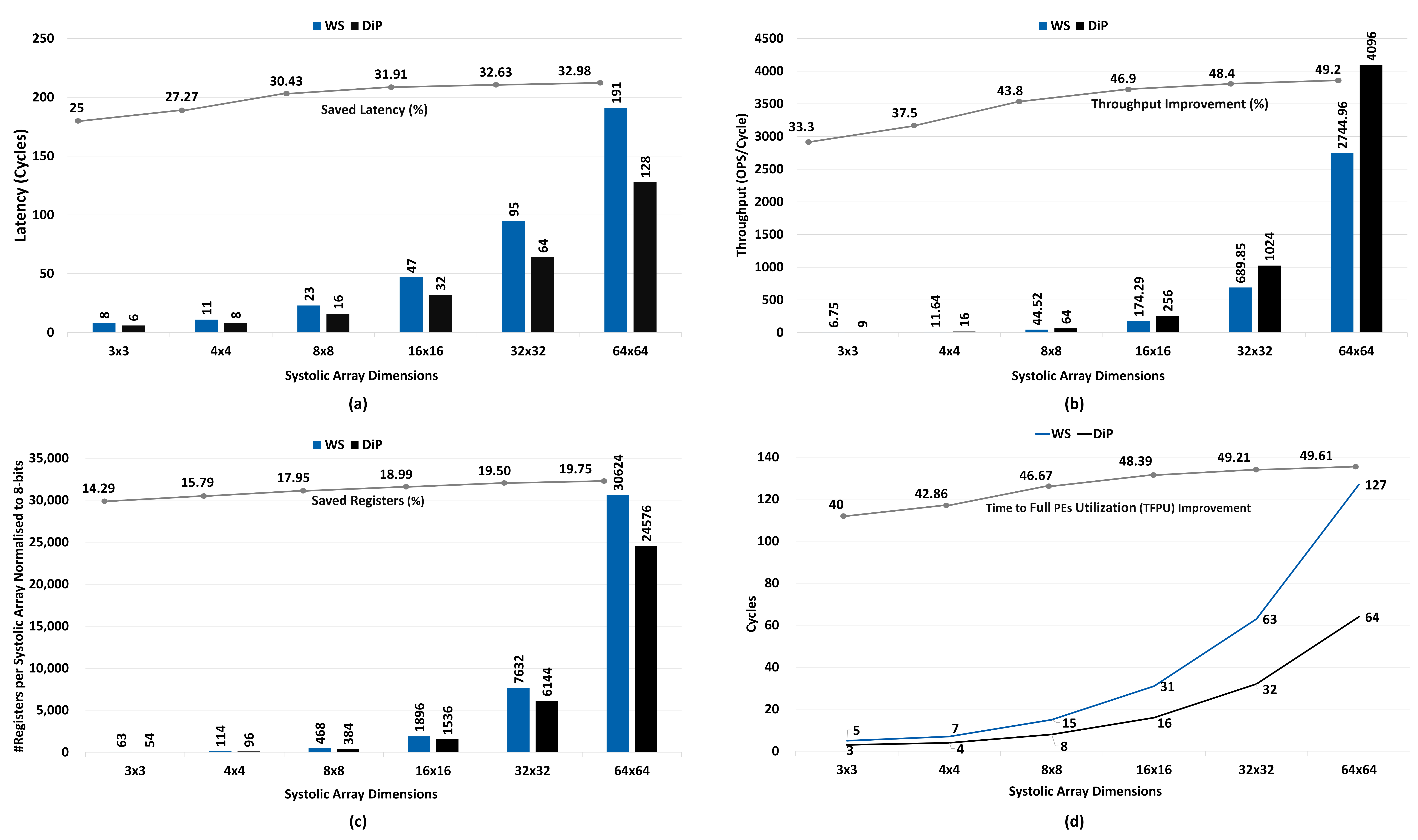} 
  \caption{Evaluation of DiP and TPU-like architecture at size 64×64 with MHA and FFN transformer workloads. The evaluation includes actual energy consumption (a, b) and latency (c, d) across various workloads dimensions of matrix multiplication (M-N-K).}
  \label{Fig.5}
\end{figure*}

\subsection{DiP Evaluation}

DiP is evaluated against a TPU-like (WS-based) architecture using transformer workloads. Nine widely used transformer models are selected to span various application domains and represent a spectrum of model sizes, from small language models (SLMs) to large language models (LLMs). These models organized in three types: Encoder-Decoder, Encoder-only, and Decoder-only. Encoder-Decoder models include Vanilla Transformer \cite{10}, T5 \cite{30}, and BART \cite{31}; Encoder-only models include BERT \cite{35}, ALBERT \cite{36}, and Transformer-XL \cite{37}; while Decoder-only models include GPT-2 \cite{32}, GPT-3 \cite{33}, and LLaMA \cite{34}. The models are selected with hyper-parameters to cover a diverse range of workloads. Sequence lengths are chosen from a range of 64 to 2048, specifically (64, 128, 256, 512, 1024, 2048). Additionally, the model's hidden size (\(d_{\text{model}}\)) varies across (512, 768, 1024, 1280, 5120), while the head size (\(d_k\)) is set to either (64, 128). Moreover, the FFN size (\(d_{\text{FFN}}\)) is configured with values of (2048, 3072, 4096, 5120).

\begin{table*}[h]
\centering
\caption{Comparison with State-of-The-Art Accelerators}
\label{table:literature_comparison}

\begin{tabular}{
>{\raggedright\arraybackslash}p{3.9cm}
>{\centering\arraybackslash}p{1.55cm}
>{\centering\arraybackslash}p{1.55cm}
>{\centering\arraybackslash}p{1.55cm}
>{\centering\arraybackslash}p{1.55cm}
>{\centering\arraybackslash}p{1.55cm}
>{\centering\arraybackslash}p{1.55cm}}

\toprule
& \textbf{\ \ DiP \newline(this work)} & \textbf{Google TPU V4i\cite{25}} & \textbf{Groq TSP \newline\cite{38}} & \textbf{\ OPTIMUS \newline\cite{39}} & \textbf{DTATrans \cite{40}} & \textbf{\ \ Eidetic \newline\cite{41}} \\

\midrule
\textbf{Architecture} &
64$\times$64, MACs &
4$\times$128$\times$128, MACs &
Tensor Processors &
SA-RCSC &
VSSA &
In-Memory Computing \\

\textbf{Design Level} &
Core-level, &
System-level, &
System-level, &
System-level, &
Core-level, &
Core-level, \\

\textbf{} &
Post-PnR &
Post-Silicon &
Post-Silicon &
Post-Syn &
Post-Syn &
Modeling \\

\textbf{Frequency (GHz)} &
1 &
1.05 &
0.9 &
0.2 &
1 &
1.5 \\

\textbf{Precision} &
INT8 &
INT8 &
INT8 &
INT16 &
Multiple &
Multiple \\

\textbf{Technology} &
22nm &
7nm &
14nm &
28nm &
40nm &
22nm \\

\textbf{Power (W)} &
0.858 &
175 &
300 &
0.732 &
0.803 &
235.7 \\

\textbf{Area (mm\textsuperscript{2})} &
1 &
400 &
725 &
5.2 &
1.49 &
266.4 \\

\textbf{Peak Throughput (TOPS)} &
8.192 &
138 &
820 &
0.5 &
1.304 &
175.5 \\

\textbf{Area Efficiency (TOPS/mm²)} &
8.192 &
0.345 &
1.131 &
0.096 &
0.979 &
0.659 \\

\textbf{Energy Efficiency (TOPS/W)} &
9.548 &
0.786 &
2.733 &
0.683 &
1.623 &
0.745 \\

\textbf{Area Efficiency (TOPS/mm\textsuperscript{2})\textsuperscript{(1)}} &
8.192 &
0.017 &
0.412 &
0.153 &
2.984 &
0.659 \\

\textbf{Energy Efficiency (TOPS/W)\textsuperscript{(1)}} &
9.548 &
0.345 &
1.823 &
0.797 &
2.470 &
0.745 \\

\bottomrule
\end{tabular}

\begin{flushleft}
\footnotesize
\textsuperscript{(1)}Power and area are normalized to 22nm using DeepScaleTool \cite{42, 43}
\end{flushleft}

\end{table*}

DiP and TPU-like architectures, each with a size of 64×64, are used for the evaluation. This architecture size aligns well with matrix tiling, as the head size for most transformer models is either 64 or 128. Cycle-accurate simulations are performed to evaluate both DiP and TPU-like implementations in terms of actual latency for each workload and energy consumption. Matrix tiling is used to process matrix multiplication workloads on DiP and TPU-like architectures by dividing the input matrices \( M_1 \) and \( M_2 \) into sub-matrices (tiles) of 64×64. By studying many transformer models, the majority of MHA and FFN workload dimensions are divisible by 64. The multiplication is performed per tile as follows; For DiP and TPU-like architectures, every tile of \( M_2 \) is loaded once and remains stationary throughout the computation for the corresponding output tile. For each tile of \( M_2 \), respective tiles from \( M_1 \) is iteratively loaded, multiplied, and saved as output psum tiles. After processing all tiles, the final output matrix is constructed by accumulating the associated psum tiles. 

Figure \ref{Fig.5} compares the energy consumption and latency of DiP and TPU-like 64x64 architectures for MHA and FFN workloads across varying dimensions (M-N-K). DiP consistently outperforms TPU-like implementation for MHA and FFN workloads. Energy improvements for MHA workloads range from 1.81x for smaller workloads to 1.25x for larger ones, while FFN workloads show a similar trend with improvements from 1.8x to 1.25x. These results highlight the energy-efficiency of DiP. Additionally, actual latency for MHA and FFN workloads demonstrates DiP's performance against TPU-like implementation, offering up to 1.49x improvements for smaller workloads, gradually reducing to approximately 1.03x for larger workloads. The breakdown of latency improvement happens for larger workloads as TPU-like architecture hides the latency associated with loading more \( M_1 \) tiles per every new \( M_2 \) tile. In contrast, for small to medium-sized workloads, TPU-like architectures incur the TFPU penalty of loading \( M_1 \) tile associated with each new \( M_2 \) tile. Additionally, TPU-like architectures still face the overhead of Input/Output synchronization FIFOs, which impact power consumption, latency, and TFPU. The evaluation highlights DiP as an energy-efficient design, making it a compelling alternative to TPU-like architectures.

\subsection{Comparison with Related Accelerators}
Table \ref{table:literature_comparison} compares DiP with Google TPU V4i\cite{25}, Groq TSP\cite{38}, OPTIMUS \cite{39}, DTATrans\cite{40}, and Eidetic\cite{41}, highlighting DiP's performance and energy efficiency for matrix multiplication acceleration. 
To ensure a fair and transparent comparison, Table \ref{table:literature_comparison} annotate whether each result corresponds to a system-level or a core-level implementation. Also, the maturity level for each design is annotated as follows: Google TPUv4i and Groq TSP represent post-silicon results; DiP is based on post-layout (post-PnR) implementation; OPTIMUS and DTATrans provide post-synthesis results; and Eidetic is based on modelling results.
Since each accelerator is implemented in different technology, the performance metrics are normalized to 22nm using DeepScaleTool\cite{42,43}. Area and energy efficiencies results are presented before and after scaling to 22nm. 
In the case of Google TPUv4i, the Matrix Multiply Units (MXUs) occupy approximately 11\% of the total chip area, yet deliver nearly the entire performance. While the power consumption of the MXUs alone is not explicitly reported, DiP provides higher area efficiency, achieving 8.192 TOPS/mm² compared to 3.14 TOPS/mm² and 0.157 TOPS/mm² for the Google TPU MXUs at 7nm and scaled 22nm, respectively.
DiP architecture features 4,096 MACs in a 64×64 configuration, operating at 1 GHz with INT8 precision using commercial 22nm technology. At a 64×64 size with 4096 PEs, DiP offers peak throughput of 8.192 TOPS, energy efficiency with 9.548 TOPS/W, and delivers an area efficiency (computational density) with 8.192 TOPS/mm². These results demonstrate DiP's capability to provide high throughput while minimizing energy consumption, showcasing its compact and optimized design. These metrics make DiP particularly well-suited for energy-efficient Transformers-based applications.

\section{Conclusion}

In this paper, a diagonal-input and permutated weight-stationary systolic array is proposed to accelerate matrix multiplication. DiP features an architecture of N×N PEs with a novel dataflow that eliminates the input and output synchronization FIFO buffers required by the conventional WS systolic array. As a result, DiP outperforms WS in terms of throughput, latency, area, and power consumption.
Additionally, the analytical models for latency, throughput, FIFO overhead, and TFPU are developed for DiP and WS architectures. The analytical model shows that DiP architecture achieves up to 50\% improvement in throughput and TFPU over WS counterparts.
Moreover, hardware design space exploration is presented for both DiP and WS architectures using commercial 22nm technology, demonstrating power consumption savings of up to 19.95\%, area savings of up to 8.12\% at 1 GHz, and energy efficiency per area improvement of up to 2.02×.
Furthermore, DiP is evaluated using various transformer workloads from widely used models such as GPT-2, GPT-3, BERT, BART, and LLaMA. DiP outperforms TPU-like architecture, achieving energy improvements ranging from 1.25× to 1.81× and latency improvements ranging from 1.03× to 1.49× across various transformer MHA and FFN workloads. DiP achieves a peak performance of 8.192 TOPS with energy efficiency of 9.548 TOPS/W. A comparison between state-of-the-art accelerators and DiP is presented, demonstrating DiP's potential as a promising architecture for energy-efficient matrix multiplication acceleration.
This paper is the foundation for DiP architecture and dataflow. Future work will focus on leveraging multiple DiP cores and deploying optimized transformers to enhance energy efficiency and performance.

\begin{IEEEbiography}
[{\includegraphics[width=1in,height=1.2in,clip,keepaspectratio]{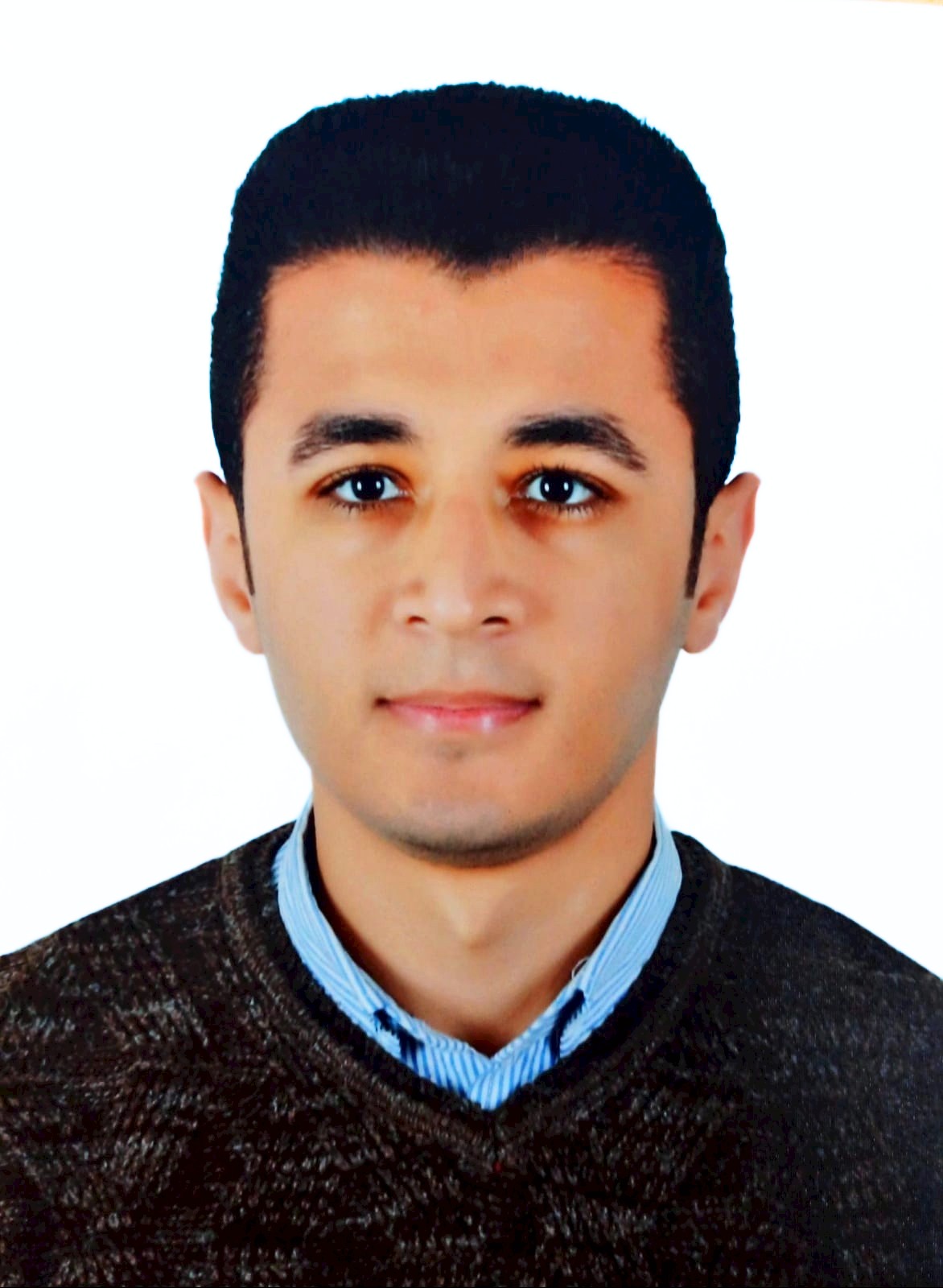}}]{Ahmed J. Abdelmaksoud}
(Member, IEEE) is currently pursuing his PhD with the Centre for Electronics Frontiers (CEF) at the University of Edinburgh, UK. He received his BSc and MSc in Electronics Engineering from Cairo University, Egypt in 2018 and 2022, receptively. Since 2018, he has been actively involved in Digital ASIC design projects across both research and industry. His professional experience includes working as a Research Associate at the System-on-Chip Center, Khalifa University, UAE; an ASIC Physical Design Engineer at Si-Vision, Egypt; and a Research Assistant at the Opto-Nano Electronics Lab, Egypt. In addition, his current research interests primarily focus on developing spatial and specialized architectures for efficient AI hardware acceleration.
\end{IEEEbiography}

\begin{IEEEbiography}
[{\includegraphics[width=1in,height=1.2in,clip,keepaspectratio]{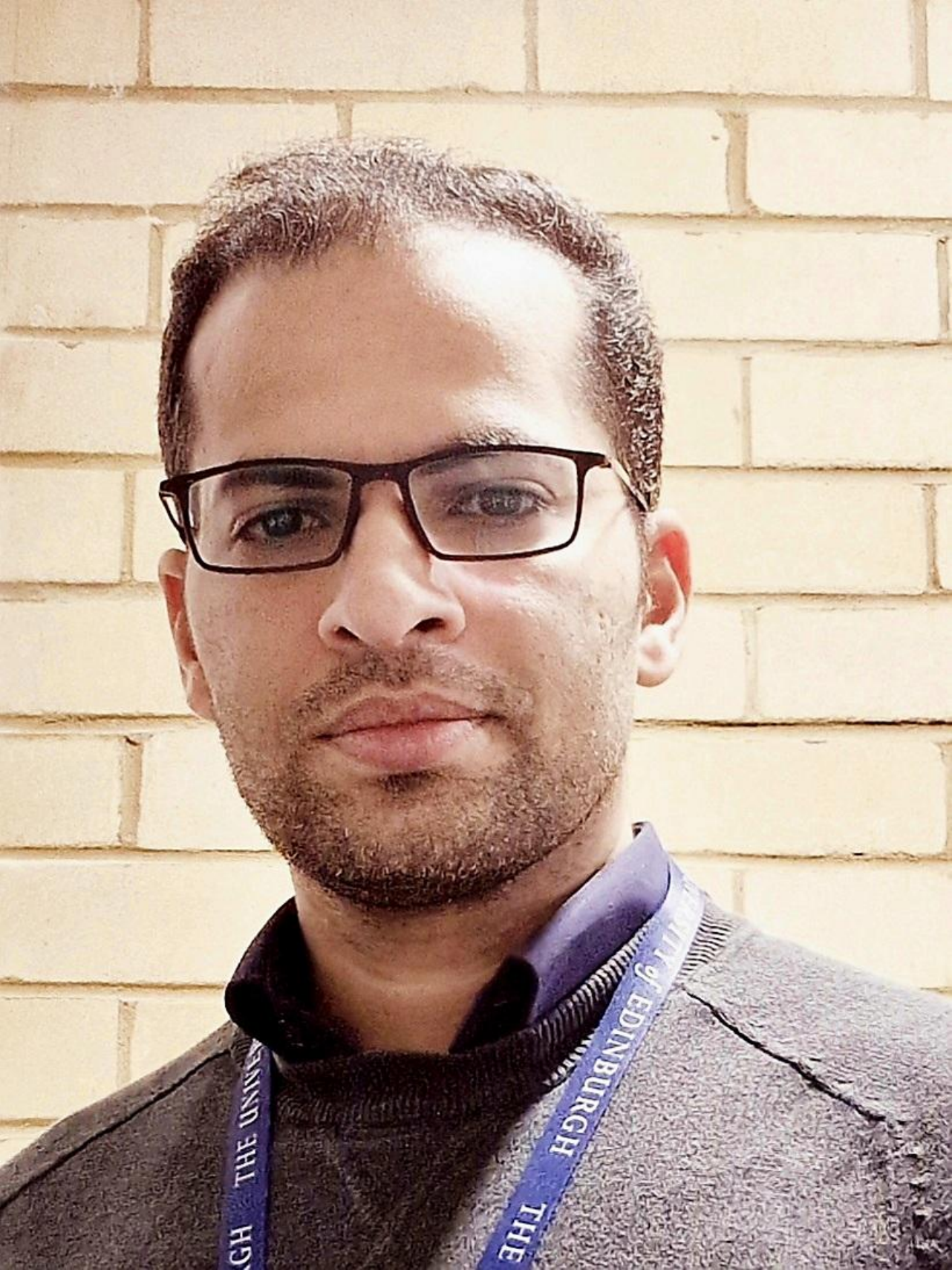}}]{Shady Agwa}
(Member, IEEE) is a Research Fellow at the Centre for Electronics Frontiers CEF, The University of Edinburgh (UK). He received his BSc and MSc degree from Assiut University (Egypt), both in Electrical Engineering. He got his PhD in Electronics Engineering from The American University in Cairo (Egypt) in 2018. Following his PhD, he joined the Computer Systems Laboratory at Cornell University (USA) as a Postdoctoral Associate for two years. In 2021, Shady joined the Centre for Electronics Frontiers at the University of Southampton (UK) as a Senior Research Fellow and then as a Research Fellow at the University of Edinburgh (UK). His research interests span across VLSI and Computer Architecture for AI using conventional and emerging technologies. His work focuses on ASIC-Driven AI Architectures with extensive expertise in In-Memory Computing, Stochastic Computing, Systolic Arrays, Beyond Von Neumann Architectures, Memories and Energy-Efficient Digital ASIC Design.
\end{IEEEbiography}

\begin{IEEEbiography}
[{\includegraphics[width=1in,height=1.2in,clip,keepaspectratio]{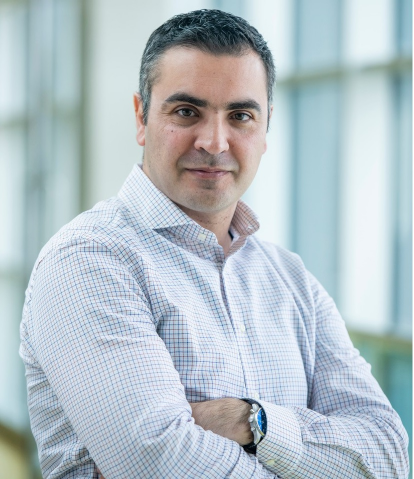}}]{Themis Prodromakis}
(Senior Member, IEEE) received the bachelor’s degree in electrical and electronic engineering from the University of Lincoln, U.K., the M.Sc. degree in microelectronics and telecommunications from the University of Liverpool, U.K., and the Ph.D. degree in electrical and electronic engineering from Imperial College London, U.K. He then held a Corrigan Fellowship in nanoscale technology and science with the Centre for Bio-Inspired Technology, Imperial College London, and a Lindemann Trust Visiting Fellowship with the Department of Electrical Engineering and Computer Sciences, University of California at Berkeley, USA. He was a Professor of nanotechnology at the University of Southampton, U.K. He holds the Regius Chair of Engineering at the University of Edinburgh and is Director of the Centre for Electronics Frontiers. He is currently a Royal Academy of Engineering Chair in emerging technologies and a Royal Society Industry Fellowship. His background is in electron devices and nanofabrication techniques. His current research interests include memristive technologies for advanced computing architectures and biomedical applications. He is a fellow of the Royal Society of Chemistry, the British Computer Society, the IET, and the Institute of Physics.
\end{IEEEbiography}

\end{document}